\documentclass[aps,pra,reprint]{revtex4-2}
\usepackage{bm}
\usepackage{amsmath}
\usepackage{amsthm}
\usepackage{amsfonts}
\usepackage{amssymb}
\usepackage{latexsym}
\usepackage[utf8]{inputenc}
\allowdisplaybreaks

\usepackage{slashed}
\usepackage{float}
\usepackage{dcolumn}
\usepackage{graphicx}

\newcolumntype{.}{D{x}{}{-1}}

\newcolumntype{w}[1]{D{.}{.}{#1}}

\usepackage{array}
\usepackage{siunitx}

\begin{document}
	
		\title{Hydrogenic rotational levels with spin-0 or spin-1/2 constituent particles}

\author{Vojt\v{e}ch Patk\'o\v{s}}
\affiliation{Faculty of Mathematics and Physics, Charles University,  Ke Karlovu 3, 121 16 Prague
	2, Czech Republic}	
	
\author{Mateusz Pa\'{n}tak}
\affiliation{Faculty of Physics, University of Warsaw,
	Pasteura 5, 02-093 Warsaw, Poland}

\author{Krzysztof Pachucki}
\affiliation{Faculty of Physics, University of Warsaw,
	Pasteura 5, 02-093 Warsaw, Poland}

\begin{abstract}
We employ nonrelativistic quantum electrodynamics with a unified description of two-body systems with spin-0 or spin-1/2 constituents, 
arbitrary masses, and arbitrary magnetic moments  in rotational states with $L>1$,
to present state-of-the-art theoretical predictions for muonic, kaonic, and antiprotonic atoms 
that have recently been measured or are targeted by upcoming experiments. 
We show that the theoretical accuracy can further be improved, opening the possibility of 
using precision spectroscopy of muonic and hadronic atoms for high-accuracy determinations of 
nuclear charge radii and nuclear electric dipole polarizabilities, and for testing the existence of hypothetical long-range hadronic interactions.
\end{abstract}
\date{\today}
\maketitle

\section{Introduction}
For an electron bound in a nucleus, we use the Dirac equation to obtain its energy levels, 
while for a scalar particle, such as a kaon or an alpha particle, we use the Klein-Gordon equation.
However, when the masses of the orbiting particle and the nucleus become comparable, 
neither the Dirac equation nor the Klein-Gordon equation is applicable. Furthermore, no differential equation can accurately describe
energy levels of such a two-body system.
In this work, we collect formulas previously derived within the framework of nonrelativistic quantum electrodynamics (NRQED), 
applicable to rotational states with $L>1$ of two-body systems consisting of spin-0 or spin-1/2 constituents 
with arbitrary masses and magnetic moments, and implement them in the PbarSpectr code \cite{patkos:25}. 
Next, we employ this code to muonic, kaonic, and antiprotonic atoms that 
have recently been measured or are the targets of ongoing and planned experiments to obtain state-of-the-art energy levels.
Finally, we argue that the theoretical accuracy can be further substantially improved, 
opening the possibility of high-precision determination of nuclear charge radii, nuclear electric dipole polarizabilities, 
and long range hadronic interactions from sufficiently accurate spectroscopic measurements.

\section{NRQED expansion of energy levels}
The spectrum of rotational hydrogenic states can be represented as a sum of all possible 
spin and angular momentum couplings,
\begin{align}
E =&\ E_\mathrm{NS} + E_\mathrm{S1}\,\vec s_1\cdot \vec L  +  E_\mathrm{S2}\,\vec s_2\cdot \vec L 
\nonumber \\ &\ + E_\mathrm{SS}\,\vec s_1\cdot\vec s_2 + E_\mathrm{LL}\,(L^i\,L^j)^{(2)}\,s_1^i\,s_2^j\,,
\end{align}
where for spin $0$ or $1/2$ particles no further spin interactions is present, and
\begin{equation}
	(L^i\,L^j)^{(2)} = \frac12\{L^i,L^j\} - \frac{\delta^{ij}}{3}\,\vec L^2 
\end{equation}
is a symmetric and traceless irreducible tensor of the second rank. 
Each component $E_X$ depends on the charge, mass, spin, and magnetic moment of constituent particles.
In the framework of NRQED, these components can be expressed as a power series in $e_1\,e_2 = -4\,\pi\,Z\,\alpha$,
namely
	\begin{equation}
		E(\alpha) = E^{(2)} + E^{(4)} + E^{(5)} + E^{(6)} + E^{(7)}+ E^{(8)}+ \ldots.
	\end{equation}
Each contribution $E^{(j)}$ is of the order $\alpha^j$, and some of them contain additional $\ln\alpha$ terms. 
Namely, $E^{(2)} = E$ is the nonrelativistic energy obtained by solving the Schr\"odinger equation with the potential $V(r)$,
which may include, for convenience, the electron vacuum polarization (EVP)
\begin{align}
V(r) = -\frac{Z\,\alpha}{r} + V^{(1)}(r) + V^{(2)}(r) + V_\mathrm{WK}(r)\,,
\end{align}
where $V^{(1)}$ is the one-loop EVP (Uehling potential), $V^{(2)}$ is the two-loop EVP, and $V_\mathrm{WK}(r)$ is the Wichman-Kroll potential.
The three-loop vacuum polarization contribution is not included in the present calculations. The uncertainty associated with its omission is estimated as $\alpha^2\,V^{(1)}$, 
in contrast to the $\alpha^2/\pi^2\,V^{(1)}$ estimate adopted in our previous work  
(see Ref. \cite{patkos:25} for further details). We employ this more conservative estimate 
because it is in better agreement with explicit calculations of three-loop vacuum polarization effects in muonic hydrogen \cite{muonic}.

The radial part of the Schr\"odinger equation takes the form
	\begin{equation}\label{01}
		P''(r) + 2\,\bigg(\mu\,\big(E-V(r)\big)-\frac{l(l+1)}{2\,r^2}\bigg)\,P(r) = 0\,,
	\end{equation}
where $\phi(r,\theta,\phi) = Y_{lm}(\theta,\phi)\,P(r)/r$. Further, $\mu$ is the reduced mass, and we assume natural units  $\hbar=c=1$.
This equation is solved numerically using the Johnson algorithm from Ref.~\cite{Johnson}, with the modification for the presence of $r\,\ln r$ terms in the small
$r$ expansion \cite{patkos:25}. This calculation presents the main part of the PbarSpectr \cite{patkos:25} code
which was applicable originally only for the case of two-body atoms with the spin $1/2$ particle and the spinless nucleus.
Here, we present the extension of PbarSpectr to the case of two-body systems 
with spin-0 or spin-1/2 constituents, arbitrary masses, and arbitrary magnetic moments, 
and the code is available as the Supplemental Material \cite{supplement}.

The main advantage of NRQED theory is that all relativistic and QED corrections can be derived
order by order in the fine structure constant $\alpha$ with controlled uncertainty for neglected terms. 
Thus, the leading relativistic $O(\alpha^2)$ correction is given by the expectation value
\begin{equation}
	E^{(4)} = \langle H^{(4)}\rangle
\end{equation}
 of the Breit Hamiltonian $H^{(4)}$ with the nonrelativistic wave function $\phi$.
For an orbiting particle with spin $s_1$, mass $m$, finite charge radius 
$r_{C1}$ and $g$-factor $g_1$, and nucleus with spin $s_2$, mass $M$,
nuclear charge radius $r_{C2}$ and $g$-factor $g_2$, the Breit Hamiltonian is \cite{Veitia, patkos:25}
\begin{widetext}
	\begin{align}\label{6}
	H^{(4)} = &\ - \frac{p^4}{8}\bigg(\frac{1}{m^3}+\frac{1}{M^3}\bigg)
	+\frac16\,\bigg(\frac{s_1\,(s_1+1)}{m^2}+\frac{s_2\,(s_2+1)}{M^2} + r_{C1}^2+r_{C2}^2\bigg)\nabla^2 V
	+\bigg[\bigg(\frac{g_1-1}{2m^2} + \frac{g_1}{2m M}\bigg)\vec L\cdot\vec s_1 \nonumber \\
	&\ + \bigg(\frac{g_2-1}{2M^2} + \frac{g_2}{2m M}\bigg) \vec L\cdot\vec s_2 \bigg] \frac{V'}{r} 
	+ \frac{1}{2mM}\bigg[\nabla^2\bigg(V - \frac14 (r V)'\bigg) + \frac{V'}{r}\vec L^2 + \bigg\{\frac{p^2}{2}\,,\,V-r V' \bigg\}\bigg] \nonumber \\
	&\ + \frac{g_1\,g_2}{6mM}\vec s_1\cdot\vec s_2\,\nabla^2 V
	-\frac{g_1\,g_2}{2(2l-1)(2l+3)mM}s_1^i s_2^j\big(L^i\,L^j\big)^{(2)}\,\bigg(\frac{V'}{r}-V''\bigg)\,,
	\end{align}
\end{widetext}
where we use the natural nuclear $g$-factors which are related 
to the standard nuclear $g_I$ factors by
\begin{equation}
	g = g_I \frac{M}{Z\,m_p}\,,
\end{equation}
and $V'$ in Eq.~(\ref{6}) stands for the derivative of the potential $V$ with respect to the radial distance $r$. 
The potential $V$ in Eq.~(\ref{6}) is the same as in the Schr\"odinger equation (\ref{01}), which enables
us to include nonperturbatively the vacuum polarization into the leading relativistic correction.

The next-order correction, which is valid for an arbitrary mass ratio, has so far been obtained without EVP and takes the form \cite{Veitia,Zatorski:22}
	\begin{align}
		E^{(5)} = &\-\frac{7\,(Z\,\alpha)^5}{3\,\pi\,\,m\,M}\,\frac{\mu^3}{l(l+1)(2l+1)n^3}
		\nonumber \\ &\
		-\frac{4\,\alpha(Z\alpha)^5\,\mu^3}{3\,\pi\,n^3}\,\biggl(\frac{1}{m}+\frac{Z}{M}\biggr)^2\,\ln k_0(n,l)\,, 
	\end{align}
where $\ln k_0(n,l)$ is the Bethe logarithm. We note that the magnetic moment anomaly is already included in $H^{(4)}$
through the exact $g$-factor. $E^{(5)} $ is numerically small for rotational levels, and the uncertainty due to the neglect of EVP is much smaller 
than the uncertainty due to unknown three-loop EVP.
	
Similarly, the higher-order $\alpha^6$ correction has been obtained without EVP
and is, in general, described by a rather lengthy expression, see Ref. \cite{Zatorski:22}.
For the $E_\mathrm{NS}$ component it takes a simple form. For two spin-less particles it is
\begin{widetext}
	  	\begin{align}
	  		E^{(6)}_{s_1=s_2=0} =& \mu (Z \alpha)^6 \biggl[ 
	  		-\frac{5}{16 n^6}+\frac{3}{2 (2 l+1) n^5}-\frac{3}{2 (2 l+1)^2 n^4}	
	  		-\frac{1}{(2 l+1)^3 n^3} \nonumber \\
	  		&\ + \frac{\mu^2}{m_1\, m_2}\bigg(\frac{3}{16\,n^6}- \frac{8\,l(l+1)-3}{2\, (2l-1)(2l+1)(2l+3)\,n^5}
	  		+ \frac{6}{(2l-1)(2l+1)(2l+3)\, n^3}\bigg)
	  		- \frac{\mu^4}{16\, m_1^2\,m_2^2\,n^6}
	  		\biggr]	\, . \label{E6spin00}
	  	\end{align} 
In the infinite-mass limit of one constituent, the result reproduces the $\alpha^6$
term of the Klein–Gordon energy expansion. Eq. (\ref{E6spin00}), however, is valid for arbitrary constituent masses. 
The analogous $E^{(6)}_\mathrm{NS}$
component for the case $s_1=0, s_2=1/2$ with $L>1$ is
\begin{align}
& E^{(6)}_{\mathrm{NS},s_1=0,s_2=1/2} - E^{(6)}_{s_1=s_2=0} \nonumber \\
= &\ \frac{\mu(Z\alpha)^6}{\Delta}
\bigg[\frac{l(l+1)}{2\,n^5}\bigg(g_2^2\frac{\mu^2}{4m_2^2} + (-2-3g_2+g_2^2)\frac{\mu^3}{2m_2^3}+\frac{\mu^4}{m_2^4}\bigg) 
+ \frac{3\lambda_0}{4\,n^4}\bigg(-g_2^2 \frac{\mu^2}{2m_2^2} + g_2 \frac{\mu^3}{m_2^3} - \frac{\mu^4}{2m_2^4}\bigg) \nonumber \\ &
+ \frac{1}{n^3}\bigg(g_2^2 \bigg(\lambda_1-\frac92\bigg)\frac{\mu^2}{4m_2^2}  + \big(3(2+5g_2-g_2^2) - 2g_2 \lambda_1\big)\frac{\mu^3}{4m_2^3} 
+(\lambda_1-9)\frac{\mu^4}{4m_2^4}\bigg) \bigg]\,,
\end{align}
and for the case $s_1= s_2=1/2$ with $L>1$ is
\begin{align}
&\ E^{(6)}_{\mathrm{NS},s_1=s_2=1/2} - E^{(6)}_{s_1=s_2=0} \nonumber \\ =&\
\frac{\mu(Z\alpha)^6}{\Delta}
\bigg[\frac{1}{n^5}\bigg(\frac{l(l+1)\,(g_1-2)}{4}\bigg(\frac{g_1+2}{2}\frac{\mu^2}{m_1^2} + (g_1-1)\frac{\mu^3}{m_1^3}\bigg)
+\frac{\mu^4}{m_1^2 m_2^2}\frac{l(l+1)}{2}
+(1\leftrightarrow2)\bigg) \nonumber \\
&\ + \frac{1}{n^4}\bigg( 
-\frac{3\lambda_0}{16} 
-\frac{3\lambda_0}{4}(g_1-2)\bigg(\frac{g_1+2}{2}\frac{\mu^2}{m_1^2}-\frac{\mu^3}{m_1^3}\bigg)
-\frac{\mu^4}{m_1^2 m_2^2}\frac{3}{8(2l+1)}\bigg(4l(l+1)
+\frac{3g_1^2 g_2^2}{16}\,
-3\bigg)+(1\leftrightarrow2)
\bigg) \nonumber \\
&\ + \frac{1}{n^3} \bigg(
\frac{\lambda_1-3}{8}
+\frac{(g_1-2)}{4}\bigg(\frac{(g_1+2)}{2}(-9+2\lambda_1)\frac{\mu^2}{m_1^2}-(-9+3g_1+2\lambda_1)\frac{\mu^3}{m_1^3}\bigg)\nonumber\\&\
+\frac{\mu^4}{m_1^2 m_2^2}\frac{1}{2}\bigg(-\frac92+\lambda_3
+\frac{g_1^2g_2^2\,\lambda_4}{16}-\lambda_4\bigg)+(1\leftrightarrow2) 
\bigg)\bigg]\,,
\end{align}
	  \end{widetext}
where
\begin{align}
	& \Delta = l(l+1)(2l-1)(2l+1)(2l+3)\,,\\
	& \lambda_0 = \frac{(2\,l-1)(2\,l+3)}{2\,l+1} \,,\\
	& \lambda_1 =\frac{3}{2l}-\frac{3}{2(l+1)} + \frac{4}{(2l+1)^2}\,\\
	&\lambda_3 = -\frac{3}{(2l-1)(2l+3)}+\frac{1}{2(2l+1)^2}\,,\\
	&\lambda_4 = -\frac{3}{4l(l+1)}-\frac{3}{(2l-1)(2l+3)}-\frac{3}{2(2l+1)^2}\,.
\end{align}

These formulas demonstrate that recoil effects depend on the spin of constituent particles.
The other components can be found in Ref.~\cite{Zatorski:22}, and they are not rewritten here due to their extensive length. 
These formulas are exact through order $\alpha^6$, 
 applicable to arbitrary mass ratios, and reduce to the corresponding Klein–Gordon or Dirac expressions in the limit of an infinitely heavy nucleus. 

One additional correction is due to the electric dipole polarizability of constituent particles
\begin{align}
E^{(6)}_\mathrm{pol} =&\ \mu\,Z^6\, \alpha^5\,
	  		\frac{2\,\mu^3\,(\alpha_{E1}+Z^{-2}\, \alpha_{E2})}{(2l-1)(2l+1)(2l+3)}
	  		\nonumber \\ 
	  		&\ \times
	  		\bigg(\frac{1}{n^5}-\frac{3}{l(l+1)\,n^3}\bigg)\,, 
			\label{E6pol}
\end{align} 
where $\alpha_{E1}$ is the static electric
dipole polarizability of the orbiting particle, while $\alpha_{E2}$ is that of the nucleus.
Nuclear static dipole polarizability is a unique property of a particular nucleus.
Nevertheless, it is useful to have a simple estimate of its magnitude.
Following Ref.~\cite{ohayon:26}, we employ the phenomenological formula
\begin{equation}\label{alphaE}
\alpha_{E2} \approx 8\,\frac{\big(\frac{A}{132}\big)^2}{\big(\frac{A}{132}\big)^\frac13 - 0.31}\,\mathrm{fm}^3,
\end{equation}
which is used to estimate the value and the associated 50\% uncertainty of the nuclear polarizability in the tables below.

Finally, higher-order QED corrections $E^{(7)}$ and $E^{(8)}$ are included in the infinite nuclear mass limit. 
$E^{(7)}$ has been calculated only for the spin $1/2$ particle (see the CODATA compilation in Ref. \cite{codata22}), 
but we use it also for the spin $0$ particle, because it should not differ significantly,
while for $E^{(8)}$ we use the Klein-Gordon and the Dirac results, correspondingly. 
Several higher-order QED corrections are omitted, such as the three-loop EVP and EVP correction to $E^{(5)}$.
Their inclusion would improve the theoretical accuracy by approximately two additional orders of
magnitude for light systems considered in this work, as illustrated in the tables below. 
Such an improvement would enable precision determinations of nuclear mean-square charge radii, 
nuclear electric dipole polarizabilities, and also  long-range hadronic interactions,
provided measurements are accurate enough.

\section{Results}
Let us now present results for systems that have been measured experimentally or studied theoretically in the literature. 
Our theoretical predictions are obtained using the updated PbarSpectr code \cite{supplement}. 
By specifying the parameters of the atomic system, such as the constituent masses, magnetic moments,
and charge radii, the code can calculate energy levels of arbitrary states of a two-body system with orbital angular
momentum $L>1$. It provides a complete breakdown of all contributions $E^{(j)}$,
together with the corresponding fine- and hyperfine-structure sublevels for chosen principal and orbital quantum numbers. Physical constants used in the calculation are summarized in Table \ref{tab:1}.
\begin{table*}
	\centering
	\caption{Table of physical constants used in the calculation. For kaon mass we use the value $m_K = 493.677(13)$~MeV from Ref.~\cite{kaonmass}.}
	\label{tab:1}
	\renewcommand{\arraystretch}{1.35}
	
	\begin{tabular}{
			c
			|S[table-format=-4.7] |S[table-format=1.10]|S[table-format=1.11]|S[table-format=-4.7]
			|S[table-format=2.5]|S[table-format=1.4]|S[table-format=-1.6]
		}
		\hline\hline
		Constant  & {$K^-$} & {$\bar{p}$\,\cite{codata22}} & {$\mu^-$\,\cite{codata22}}& {${}^{20}$Ne}
		& {${}^{19}$F} & {${}^{28}$Si} & {${}^{29}$Si}
		\\
		
		\hline
       
        $r_C$ [fm] \cite{radii}
		& {0.599}
		& {0.84075}
		& 
		& {3.0055} 
		& {2.8976}
		& {3.1224}
		& {3.1176}
		\\
		
		$g$
		&
		& 5.5856946893
		& 2.00233184123
		&
		& 11.01340
		&
		& -2.280436
		\\
		
		spin
		& {0}
		& {1/2}
		& {1/2}
		& {0}
		& {1/2}
		& {0}
		& {1/2}
		\\
		
		mass ratio $m/m_e$
		& {966.103}
		& {1836.152673426}
		& {206.768\,2827(46)}
		&{36433.99594496}
		& {34622.97577399}
		& {50984.8327636}
		& {52806.93399986}
		\\

		\hline\hline
	\end{tabular}
\end{table*}

\subsection{$\mu^{20}$Ne}
The first example is $\mu^{20}$Ne, which is planned to be measured at J-PARC \cite{prl:23:muAtom}.
Table \ref{muonic:1} presents values for individual contributions to the energy of the specified atomic  level.
We observe an excellent convergence in powers of $\alpha$, so that
the higher-order terms $E^{(7)}$ and $E^{(8)}$ for this particular case are too small to be included.
The dominant (first) uncertainty comes from the three-loop EVP. 
It has recently been calculated \cite{onishschenko,adkins:25},
but its inclusion here requires significant work and has therefore been deferred to a future version of PbarSpectr. 
The next (second) uncertainty is due to the nuclear static electric dipole polarizability, 
and is estimated by 50\% of Eq.~(\ref{alphaE}). 
Table \ref{muonic:1} illustrates the capability of the NRQED approach to provide highly accurate predictions for rotational states of exotic atoms containing a muon (as well as pion, kaon, or antiproton) bound to an arbitrary nucleus.

The comparison to former calculations from Ref.~\cite{prl:23:muAtom} is presented in Table \ref{muonic:2}.
The results of those calculations were presented without any uncertainties, and we observe that their last two digits are redundant.
Clearly, the use of the Dirac equation with reduced mass has limited applicability.
If the theoretical results for transitions presented in Table \ref{muonic:2} are compared with
the corresponding experimental measurements, they can be used to obtain the electric dipole polarizability of the $^{20}$Ne nucleus.
For that, however, they have to be measured with at least 9-digit precision (of
the 6 keV transition energy). 
This is probably not possible with the current technology, but at least 
the theoretical predictions can be obtained accurately enough by including the three-loop EVP
potential $V^{(3)}$.
	
	\begin{table*}[!]
		\centering
		\caption{Contributions to $E_\mathrm{NS}$ and $E_\mathrm{S1}$ coefficients
			 for energy levels in $\mu^{20}$Ne, in eV.
			 $\delta_\mathrm{EVP}$ denotes the part of the energy $E^{(2)}$ which comes
			 from the vacuum polarization and is used for the estimate of the first uncertainty, 
			 associated with the uncalculated three-loop vacuum polarization correction. 
		         The second uncertainty originates from the static electric dipole polarizability of the nucleus, for which only a rough estimate is currently available.}
		\label{muonic:1}
		\renewcommand{\arraystretch}{1.25}
		
		\begin{tabular}{
				c
				|S[table-format=-5.8]
				|S[table-format=1.5]
				|S[table-format=-5.9]
				|S[table-format=1.4]
				|S[table-format=-5.9]
				|S[table-format=1.5]
				|S[table-format=-5.8]
				|S[table-format=1.4]
			}
			\hline\hline
			
			&
			\multicolumn{2}{c|}{$n=5,\; l=4$}
			&
			\multicolumn{2}{c|}{$n=4,\; l=3$}
			&
			\multicolumn{2}{c|}{$n=5,\; l=3$}
			&
			\multicolumn{2}{c}{$n=4,\; l=2$}
			\\
			\hline
			Term
			& {$E_\mathrm{NS}$} & {$E_\mathrm{S1}$}
			& {$E_\mathrm{NS}$} & {$E_\mathrm{S1}$}
			& {$E_\mathrm{NS}$} & {$E_\mathrm{S1}$}
			& {$E_\mathrm{NS}$} & {$E_\mathrm{S1}$}
			\\
			
			\hline
			
			$E^{(2)}$
			& -11189.86264(2)
			& 
			& -17486.2598(2)
			& 
			& -11190.81993(8)
			&
			&-17490.0383(4)
			&
			\\
			
			$\delta_\mathrm{EVP}$
			& -0.46093(2)
			& 
			& -2.8197(2)
			& 
			&-1.41822(8)
			&
			&-6.5982(4)
			&
			\\
			
			$E^{(4)}$
			& -0.86438
			& 0.13278
			& -2.2971
			& 0.5562
			&-1.62248
			&0.28480
			&-4.9680
			&1.5607
			\\
			
			$E^{(5)}$
			& 0.00005
			&
			&0.0002
			&
			&0.00015
			&
			&0.0009
			&
			\\
			
			$E^{(6)}$
			& -0.00012
			& 0.00003
			&-0.0006
			&0.0002
			&-0.00041
			&0.00014
			&-0.0026
			&0.0016
			\\
						
			$E_\mathrm{FNS}$
			& -0.00004
			&
			&-0.0004
			&
			&-0.00020
			&
			&-0.0016
			&
			\\
			
			$E_\mathrm{pol}$
			& -0.000007(3)
			&
			& -0.000050(25)
			&
			& -0.000029(14)
			&
			& -0.00035(18)
			&
			\\
			\hline
			
			Sum
			& {$-11190.72715(2)(1)$}
			& 0.13281
			&{$-17488.5577(2)(1)$}
			&0.5565
			&{$-11192.44289(8)(1)$}
			&0.28494
			&{$-17495.0099(4)(2)$}
			&1.5623
			\\
			
			\hline\hline
		\end{tabular}
		\end{table*}

	\begin{table}[!ht]
	\centering
	\caption{Comparison of theoretical results for transitions in $\mu^{20}$Ne, in eV. 
	Transition energy $\Delta E_\mathrm{theo}$ is our theoretical result without the nuclear polarizability correction,
	which is presented separately. The polarizability contribution to the transition energy is obtained
	from the difference of the unrounded level polarizability corrections.}
	\label{muonic:2}
	\renewcommand{\arraystretch}{1.35}
	
	\begin{tabular}{
			c
			|S[table-format=-4.8]
			|S[table-format=-0.8]
			|S[table-format=-2.5]
		}
		\hline\hline
		Transition
		& {$\Delta E_\mathrm{theo}$} & {$\Delta E_\mathrm{pol}$} & {$\Delta E_\mathrm{theo}$ Ref.~\cite{prl:23:muAtom}}
		\\
		
		\hline
		
		$5g_{9/2}-4f_{7/2}$
		& {6297.2615(2)}
		& 0.00004(2)
		& 6297.26191 
		\\
		
		$5g_{7/2}-4f_{7/2}$
		& {6296.6638(2)}
		& 0.00004(2)
		& 6296.66427 
		\\
		
		$5g_{7/2}-4f_{5/2}$
		& {6298.6115(2)}
		& 0.00004(2)
		& 6298.61192
		\\
		
		$5f_{7/2}-4d_{5/2}$
		& {6301.4319(4)}
		& 0.00032(16)
		& 6301.43265
		\\
		
		$5f_{5/2}-4d_{5/2}$
		& {6300.4346(4)}
		& 0.00032(16)
		& 6300.43536
		\\
		
		$5f_{5/2}-4d_{3/2}$
		& {6304.3402(4)}
		& 0.00032(16)
		& 6304.34099
		\\
		
		\hline\hline
	\end{tabular}
	
\end{table}

\subsection{$\bar p\, ^{28}$Si and $\bar p\, ^{29}$Si} 
The next example is antiprotonic Si, which is currently investigated by the PAX collaboration \cite{PAX}. 
Apart from having a much larger magnetic moment anomaly, the antiproton is almost 10 times heavier than muon.
One therefore expects the NRQED approach, which is exact with respect to the mass ratio, to be
particularly well suited for such systems. Indeed, for the transition between $6h$ and $5g$ levels
we observe excellent convergence of the expansion in powers of $\alpha$.
We consider both stable silicon isotopes $^{28}$Si and $^{29}$Si, with the nuclear spin $0$ and $1/2$, respectively.
Among all spin couplings, the antiproton spin-orbit interaction is the largest one, while the nuclear spin-orbit interaction is about 60 times smaller.
Remaining spin couplings are smaller than the uncertainty due to the nuclear polarizability.
The results presented in Table \ref{antiprotonic} lead to the  transition energy in $\bar{p}\,{}^{28}$Si
\begin{align}
\Delta E(6h_{11/2}-5g_{9/2},{}^{28}\mathrm{Si})=58029.57(3)(13)\;\mathrm{eV}, 
\end{align}
where the first uncertainty is due to the higher-order QED, and the second is due to static dipole polarizability of the nucleus and antiproton.
A similar result for $\bar{p}\,{}^{29}$Si, averaged over all hyperfine levels, is
\begin{align}
\Delta E(6h_{11/2}-5g_{9/2},{}^{29}\mathrm{Si})=58099.44(3)(14)\;\mathrm{eV}.
\end{align}
The resulting isotope shift is
\begin{align}
\Delta E(6h_{11/2}-5g_{9/2}, {}^{29}\mathrm{Si} - {}^{28}\mathrm{Si} ) = 69.87(14)\;\mathrm{eV},
\end{align}
where the uncertainty due to higher-order QED is canceled, and we estimate 
the uncertainty due to nuclear polarizability by taking the larger one from the individual values.
This isotope shift thus represents an excellent test of the NRQED approach.
	\begin{table*}[!ht]
	\centering
	\caption{Contributions to coefficients $E_\mathrm{NS}$, $E_\mathrm{S1}$, $E_\mathrm{S2}$, 
	             $E_\mathrm{SS}$, and $E_\mathrm{LL}$ for $6h$ and $5g$ states in $\bar{p}\,{}^{28}$Si and $\bar{p}\,{}^{29}$Si, in eV.} 
	\renewcommand{\arraystretch}{1.35}
	\label{antiprotonic}
	
	\begin{tabular}{
			c
			|S[table-format=-5.5]|S[table-format=-1.4]
			|S[table-format=-5.5]|S[table-format=-1.4]|S[table-format=-1.4]|S[table-format=-1.4]|S[table-format=-1.4]
		}
		\hline\hline
		&
		\multicolumn{2}{c|}{${}^{28}$Si$\,(n=6, l=5)$} & 
		\multicolumn{5}{c}{${}^{29}$Si$\,(n=6, l=5)$}
		\\ \hline
		Term 
		& {$E_\mathrm{NS}$} & {$E_\mathrm{S1}$} &
		{$E_\mathrm{NS}$} & {$E_\mathrm{S1}$} & {$E_\mathrm{S2}$} & {$E_\mathrm{SS}$} & {$E_\mathrm{LL}$}
		\\
		
		\hline
		
		$E^{(2)}$
		& -131556.18(1)
		&
		& -131714.48(1)
		&
		&
		&
		&
		\\
		
		$	\delta_\mathrm{EVP}$
		& -270.45(1)
		&
		& -271.09(1)
		&
		&
		&
		&
		\\
		
		$ E^{(4)}$
		& -13.4057
		& 6.2259
		& -13.4115
		& 6.2393
		& -0.1087
		& 0.0003
		& 0.0148
		\\
		
		$ E^{(5)}$
		& 0.0000
		&
		& 0.0001
		&
		&
		&
		&
		\\
		
		$ E^{(6)}$
		& -0.0071
		& 0.0027
		& -0.0071
		& 0.0028
		& -0.0001
		& 0.0002
		& 0.0000
		\\
		
		$ E_\mathrm{FNS}$
		& -0.1494
		&
		& -0.1495
		&
		&
		&
		&
		\\
		
		$E_\mathrm{pol}$
		& -0.05(3)
		&
		& -0.06(3)
		&
		&
		&
		&
		\\
		
		\hline
		
		Sum
		& {-131569.80(1)(3)}
		& 6.23
		& {-131728.11(1)(3)}
		& 6.24
		& -0.11
		& 0.00
		& 0.01
		\\
\hline
		&
\multicolumn{2}{c|}{${}^{28}$Si$\,(n=5, l=4)$} & 
\multicolumn{5}{c}{${}^{29}$Si$\,(n=5, l=4)$}
\\ \hline

Term 
& {$E_\mathrm{NS}$} & {$E_\mathrm{S1}$} &
{$E_\mathrm{NS}$} & {$E_\mathrm{S1}$} & {$E_\mathrm{S2}$} & {$E_\mathrm{SS}$} & {$E_\mathrm{LL}$}
\\

\hline

$E^{(2)}$
& -189593.11(3)
&
& -189821.31(3)
&
&
&
&
\\

$	\delta_\mathrm{EVP}$
& -541.65(3)
&
& -542.82(3)
&
&
&
&
\\

$ E^{(4)}$
& -29.4724
& 19.7693
& -29.4865
& 19.8119
& -0.3451
& 0.0009
& 0.0716
\\

$ E^{(5)}$
& 0.0003
&
& 0.0003
&
&
&
&
\\

$ E^{(6)}$
& -0.0310
& 0.0142
& -0.0311
& 0.0142
& -0.0003
& 0.0008
& 0.0003
\\

$ E_\mathrm{FNS}$
& -0.4904
&
& -0.4908
&
&
&
&
\\

$E_\mathrm{pol}$
& -0.26(13)
&
& -0.28(14)
&
&
&
&
\\

\hline

Sum
& {-189623.36(3)(13)}
& 19.78
& {-189851.59(3)(14)}
& 19.83
& -0.35
& 0.00
& 0.07
\\

		\hline\hline
	\end{tabular}
	
\end{table*}

\subsection{$K^{19}$F}
The first example of the system with a scalar orbiting particle is the kaonic fluorine, presented in Table \ref{kaonic:1}. 
The nucleus has spin $1/2$, which leads to hyperfine splitting. However, we present results only for the difference of centroid energies, 
which were also calculated by Paul Indelicato in Ref.~\cite{arxiv:26:KF}. 
For the highest transition $6h - 5g$ we observe only a small difference of $0.01$ eV, but the difference grows with decreasing principal quantum number,
reaching $0.03$ eV for $6h - 4f$ and $0.15$ eV for $4f - 3d$. In Table \ref{kaonic:1} we do not account for the strong interaction shift,
which was estimated for the lowest $4f - 3d$ transition to be about -2.0 eV, and is included in the results of Ref.~\cite{arxiv:26:KF}.  
The relatively large uncertainty associated with the limited precision of the kaon mass suggests that these
 transitions could be used for its determination, provided that the experimental precision is 
 improved to the level of approximately 0.1 eV. The current experimental accuracy 
 is of the order of 10 eV.
Once the uncertainty of the kaon mass is reduced, spectroscopic measurements may become 
sensitive to subtler effects such as the long-range hadronic interactions.

	\begin{table*}[!ht]
	\centering
	\caption{Comparison of theoretical results for centroid transitions in $K^{19}$F , in eV.
		The uncertainty of the QED result is due to the omitted 3-loop vacuum polarization.
		We did not take into account strong interaction shift of $-2$~eV for
		$4f-3d$ transition \cite{arxiv:26:KF}.
		Uncertainty due to the kaon mass can be estimated as approximately $2.6\cdot10^{-5}\times \Delta E_\mathrm{theo}$.}
	\label{kaonic:1}
	\renewcommand{\arraystretch}{1.35}
	
	\begin{tabular}{
			c
			|S[table-format=-5.7]
			|S[table-format=1.6]
			|S[table-format=-0.3]
			|S[table-format=-7.6]
		}
		\hline\hline
		Transition
		& {$\Delta E_\mathrm{theo}$} & {$\Delta E_\mathrm{pol}$} & {$\Delta E_\mathrm{theo}$ Ref.~\cite{arxiv:26:KF}}
		& {$\Delta E_\mathrm{exp}$ Ref.~\cite{arxiv:26:KF}}
		\\
		
		\hline
		
		$6h-5g$ 
		& {12684.703(2)}
		&  0.001(1)
		& 12684.69${}^{+0.42}_{-0.43}$
		& {$12673.9\pm5.9\pm7.1$}
		\\
		
		$7i-5g$
		& {20327.508(2)}
		& 0.002(1)
		& 20327.49${}^{+0.59}_{-0.64}$
		& {$20298.8\pm16.3\pm6.0$}
		\\
		
		$5g-4f$
		& {23377.397(6)}
		& 0.012(7)
		& 23377.38${}^{+1.06}_{-1.06}$
		& {$23383.2\pm4.7\pm5.5$}
		\\
		
		$8k-5g$
		& {25285.577(2)}
		& 0.002(1)
		& 25285.56${}^{+0.70}_{-0.85}$
		& {$25279.6\pm36.2\pm5.0$}
		\\
		
		$6h-4f$
		& {36062.100(6)}
		& 0.013(7)
		& 36062.07${}^{+1.27}_{-1.28}$
		& {$36073.3\pm30.7\pm30.0 $}
		\\
		
		$4f-3d$
		& {50592.63(2)}
		& 0.19(10)
		& 50590.48${}^{+4.36}_{-4.36}$
		& {$50586.7\pm24.3\pm23.0$}
		\\

		\hline\hline
	\end{tabular}
	
\end{table*}

\subsection{$K^{20}$Ne}
The final system considered in this work is kaonic neon, a two-body bound system consisting of two spin-0 particles.
For the highest transition considered, $9l-8k$, our result agrees with previous theoretical predictions to within $0.01$ eV.
The discrepancy increases gradually with decreasing principal quantum number, reaching approximately $0.07$ eV for the $6h-5g$ transition.
The higher-order relativistic and QED corrections
$E^{(7)}$ and $E^{(8)}$ are too small ($\sim 10^{-6}$~eV) to account for this difference. 
We conclude that, once relativistic effects become significant, the exact dependence of 
relativistic corrections on the masses of both constituents becomes important. 
Finite nuclear mass effects cannot be fully reproduced within the reduced-mass approximation 
and require an explicit two-body treatment rather than a simple reduced-mass substitution in the relativistic corrections.
	\begin{table*}[!ht]
	\centering
	\caption{Comparison of theoretical results for centroid transitions in $K^{20}$Ne, in eV. 
	Uncertainty due to the kaon mass can be estimated as approximately $2.6\cdot10^{-5}\times \Delta E_\mathrm{theo}$.}
	
	\renewcommand{\arraystretch}{1.35}
	
	\begin{tabular}{
			c
			|S[table-format=-5.7]
			|S[table-format=1.6]
			|S[table-format=-4.2]
			|S[table-format=-5.5]
		}
		\hline\hline
		Transition
		& {$\Delta E_\mathrm{theo}$} & {$\Delta E_\mathrm{pol}$} & {$\Delta E_\mathrm{theo}$ Ref. \cite{26:KNe}}
		& {$\Delta E_\mathrm{exp}$ Ref. \cite{26:KNe,25:KNe}}
		\\
		
		\hline
		
		$9l-8k$ 
		& 4201.4519(2)
		& 0.0000
		& 4201.45
		& {$4206.97\pm3.43\pm2.00$}
		\\
		
		$8k-7i$
		& 6130.3227(5)
		& 0.0001(1)
		& 6130.31 
		& {$6130.57\pm0.65\pm1.50$}
		\\	
		
		$7i-6h$
		& 9450.306(1)
		& 0.0005(3)
		& 9450.28 
		& {$9450.23\pm0.37\pm1.50$}
		\\	
		
		$6h-5g$
		& 15685.462(3)
		& 0.002(2)
		& 15685.39 
		& {$15673.30\pm0.52\pm9.00$}
		\\			
		
		\hline\hline
	\end{tabular}
\end{table*}

\section{Summary}

We have employed the NRQED approach and implemented formulas in the PbarSpectr code 
for $E^{(4)}$, $E^{(5)}$, and $E^{(6)}$, with nonperturbative inclusion of EVP in $E^{(4)}$, 
in two-body systems containing spin-0 and spin-1/2 particles. 
The obtained results were compared with previous relativistic calculations based on the Klein-Gordon and Dirac equations. 
Since the NRQED formalism accounts for the exact mass dependence of relativistic and QED corrections,
we observe small differences of the order of $\sim 10^{-7}$ with respect to these former approximate calculations.

The theoretical predictions can be further improved by about two orders of magnitude 
through the incorporation of three-loop EVP and one-loop EVP with $E^{(5)}$. 
Such an improvement would open the possibility of determining nuclear 
mean-square charge radii and nuclear polarizabilities from precision 
spectroscopy of muonic and antiprotonic atoms.
In the case of pionic and kaonic atoms,
the main uncertainty of theoretical predictions comes from the masses of the orbiting scalar particles,
so these masses are the first to be determined more accurately from atomic spectroscopy.

The NRQED approach for heavy muonic and antiprotonic atoms
can be supplemented by exact-in-$Z\alpha$ formulas for expansion coefficients in the mass ratio within HPQED formalism \cite{hpqed}.
This will enable high-precision tests of QED in heavy hydrogen-like muonic and antiprotonic atoms 
and will provide a sensitive probe for hypothetical long-range interactions beyond the Standard Model.

\acknowledgments
We gratefully acknowledge many interesting comments and suggestions from Ben Ohayon.

\end{document}